\newdimen\offdimen

\def\offset#1#2{\offdimen #1
   \noindent \hangindent \offdimen
   \hbox to \offdimen{#2\hfil}\ignorespaces}

\input amstex
\documentstyle{amsppt}
\magnification=1200
\tolerance 1000
\hsize 16.4 truecm
\vsize 22.9 truecm

\global\topskip=0pt
\baselineskip 18pt

\def\A{\Cal A}
\def\C{\Cal C}

\def\F{\Cal F}
\def\P{\Cal P}
\def\g{\frak g}
\def\V{\Cal V}
\def\h{\frak h}
\def\L{\Cal L}
\def\n{\frak n}
\def\b{\frak b}

\def\M{\frak M}

\def\B{\bold B}
\def\G{\bold G}
\def\N{\bold N}
\def\PP{\bold P}
\def\T{\bold T}

\topmatter
\title
NON-COMMUTATIVE AND COMMUTATIVE INTEGRABILITY OF GENERIC TODA FLOWS
IN SIMPLE LIE ALGEBRAS
\endtitle

\author
M.I. Gekhtman$^\star ${}$^\dagger$ and M.Z. Shapiro$^{\star\star}$ \\ 
{\eightrm $^\star$ Department of Mathematics,\\
 The University of Michigan,\\
Ann Arbor, USA \\
e-mail: gekhtman\@math.lsa.umich.edu\\
\\
$^{\star\star}$Matematiska Institutionen, KTH \\
Stockholm, Sweden\\
e-mail:mshapiro\@math.kth.se}
\endauthor

\leftheadtext{M.I. Gekhtman, M.Z. Shapiro}
\rightheadtext{Generic Toda Flows}

\thanks
{}$^\dagger$Research partially supported by
AFOSR grant F49620-96-1-0100
\endthanks



\endtopmatter

\document

\head 1. Introduction
\endhead

{\bf 1.1.} The Toda lattice equation introduced by Toda as a Hamilton equation
describing the motion of the system of particles on the line
with an exponential interaction between closest neighbours gave
rise to numerous important generalizations and helped to discover
many of the exciting phenomena in the theory of integrable equations.
In Flaschka's variables [F] the finite non-periodic Toda lattice describes
an isospectral evolution on the set of tri-diagonal matrices in $sl(n)$.
It was explicitly solved and shown to be completely integrable by
Moser [Mo] . 

In his paper [K1], Kostant comprehensively studied
the generalization of Toda lattice that evolves on the set
of ``tri-diagonal'' elements of a semisimple Lie algebra $\g$ which 
also turned out
to be completely integrable with Poisson commuting integrals being 
provided by the Chevalley invariants of the algebra. Moreover, in this
paper, as well as in the works by Ol'shanetsky and Perelomov [OP],
Reyman and Semenov-Tian-Shansky [RSTS1], Symes [Sy], the method of
the explicit integration of the Toda equations was extended to the
case when evolution takes place on the dual space of the Borel 
subalgebra of $\g$ . This space is foliated into 
symplectic leaves of different dimensions and the natural question
is what can be said about the Liouville complete integrability of the
Toda flows on each of these leaves. In the particular case of generic
symplectic leaves in $sl(n)$ the complete integrability was proved
by Deift, Li, Nanda and Tomei [DLNT]. 

This paper was motivated by the work  [DLNT] and its Lie 
algebraic interpretation proposed in [S1], [S2], [EFS].
Our  main result is the following

\proclaim{Theorem 1.1} The Toda flows on generic coadjoint orbits 
in simple Lie algebras are completely integrable.
\endproclaim

{\bf 1.2.} Recall that there are several versions of the Toda equation in
$sl(n)$ corresponding to different realizations of the dual 
$\b_+^*$ of a subalgebra $\b_+$ of upper triangular matrices. 
In particular, we can use the trace form to identify  $\b^*$ with either
linear subspace $S$ of symmetric matrices or affine subspace 
$\epsilon + \b_-$ of lower Hessenberg matrices, where
$\b_-$ is a  subalgebra of lower triangular matrices and 
$\epsilon=(\epsilon_{jk})= (\delta_{j+1,k})$ . The Toda equation
associated with the first realization is called symmetric. Following
[EFS],  we  call the Toda
equation on  lower Hessenberg matrices {\it the Kostant-Toda equation}.

Both the symmetric Toda equation and 
the Kostant-Toda equation can be written in the Lax form 
$$ \dot X = [X, B(X)]\ , \eqno(1.1)$$
where $B(X)$ is a projection of $X$ along the subspace of upper triangular
matrices onto a skew-symmetric  matrix if $X$ is symmetric or 
lower triangular matrix if $X$ is lower Hessenberg. In both cases, equation
(1.1) is Hamilton equation with a Hamiltonian given by 
$H(X)={1\over 2} Tr(X^2)$
and a Poisson structure obtained as a pull-back from the standard Lie-Poisson
structure on the dual of the algebra of upper triangular matrices. The Toda
equation defines a flow on the symplectic leaf that contains initial data
$X_0$ and coincide with the orbit through $X_0$ of the coadjoint action
of the upper triangular group.

When restricted to tridiagonal matrices, equation (1.1), called the Toda
lattice, is completely integrable. The maximal Poisson commutative family
is provided then by eigenvalues or, equivalently, traces of powers of $X$ .

The set of tridiagonal matrices in $S$ or $\epsilon + \b_-$ can be viewed
as the minimal indecomposable symplectic leaf. 
Deift, Li, Nanda and Tomei [DLNT] addressed the ``maximal''  
case and proved
that the symmetric Toda equation
is completely integrable on symplectic leaves containing sufficiently
generic full matrices $X$ or, in other words,
on generic coadjoint orbits. They  considered
the minors
$$P_k(\lambda) = \det (X - \lambda \operatorname{Id})_k
, k=0,\dots, [{{(n-1)} \over 2}] \ ,\eqno(1.2)$$
where $(M)_k$ is obtained from
the matrix $M$ obtained by deleting first
$k$ rows and last $k$ columns. If $X$ is {\it generic}, i.e. if all
lower left $k\times k$ minors of $X$ ( $k=1,\dots,[n/2]$ ) are nonzero, then
$$P_k(\lambda) = 
E_{0k} ( \lambda^{n-2k} + I_{1k}\lambda^{n-2k} \cdots +  I_{n-2k,k})\ .
\eqno(1.3) $$
Functions $ I_{jk}, \ 
j = 2, \dots, n-2k$ are invariant under the adjoint action of the parabolic
subgroup $\PP_k \subset SL(n)$ of matrices whose strictly lower triangular 
parts have all zero entries in the first $k$ columns and last $k$ rows .
This observation was used in [DLNT] in order to combine the 
Adler-Kostant-Symes theorem with Thimm's method [T] of construction of
involutive families using nested chains of parabolic subalgebras and to
show that $ I_{jk}, \ k=0,\dots, [{{(n-1)} \over 2}], \ 
j = 2, \dots, n-2k$
form a maximal involutive family of integrals
of the Toda flow, while $ I_{1k}, k=0,\dots, [{{(n-1)} \over 2}]$ are 
coadjoint invariants that completely determine the symplectic leaf.

Definitions and statements of the previous paragraph remain valid without
any changes for the case of the Toda flows on generic coadjoint orbits
in $\epsilon + \b_-$ even though full symmetric and Kostant-Toda flows
are not isomorphic and exhibit different dynamical behaviour. However,
in what concerns  a complete integrability of the Toda equation, it is
lower triangular part of $X$ that really matters. Bearing this in mind, 
we shall restrict ourselves to the Kostant realization of the Toda flows.

It was observed by Singer [S1,S2] that if one represents $(X)_k$ as a 
$2\times 2$ block matrix $\pmatrix U_k & X_k\\ Y_k & Z_k \endpmatrix$,  where
$X_k$ and $Y_k$ are $(n-2k)\times (n-2k)$ and $k\times k$ blocks resp., 
then $ I_{jk}, \ j = 2, \dots, n-2k$ are coefficients of the characteristic
polynomial $\det ( \lambda - \phi_k )$ of {\it the Schur complement}
$$\phi_k (X) = X_k - U_k Y^{-1}_k Z_k \eqno(1.4)$$
of $X_k$ in $(X)_k$ . Moreover, it was shown in [S1,S2] that the 
{\it $k$-chop} $\phi_k$
considered as a map from $sl(n)$ into $sl(n-2k)$ is Poisson w.r.t. both
the Lie-Poisson bracket on and the {\it $R$-matrix bracket} whose
restriction to $\epsilon + \b_-$ provides a Poisson structure for the
Toda flow (for definitions, see section 2.3 below) . In particular, this
gives another proof of the involutivity of $ I_{jk}$ .

In [A] it was shown that generic matrix $X$ can be conjugated by an element
of the upper triangular group to the following form
$$
Ad_{b_X} X=\pmatrix
\varkappa_1 & * &* &\dots &\dots &* \\
0&\varkappa_2&* &\dots &\dots& *\\
\vdots&\vdots &\ddots&\ddots&\ddots&\vdots\\
0 &0  &  &\ddots &\ddots &\vdots \\
0 &1 &0 &\dots &\varkappa_2 &*\\
1 &0 &0   &\dots  &0  &\varkappa_1
\endpmatrix
\ .
\eqno(1.5)
$$
Here the only nonzero entries in the lower triangular part of $Ad_{b_X} X$
are units on the anti-diagonal, while the diagonal part of  $Ad_{b_X} X$
is symmetric w.r.t the anti-diagonal. The element $b_X$ is defined uniquely
up to a right multiplication by an  invertible diagonal matrix symmetric w.r.t
the anti-diagonal. This ambiguity does not affect values of 
$ \varkappa_1,\varkappa_2,\dots $ and thus, $ \varkappa_1,\varkappa_2,\dots $
are coadjoint invariants and (1.5) can be considered as a normal form
of a generic coadjoint orbit through $X$ . Note that the normal form (1.5)
can be obtained in several similar steps, the first one being a conjugation
with an upper triangular matrix 
$$
\Gamma =\pmatrix
\gamma_{11} & \gamma_{12} &\dots &\dots &\dots &\gamma_{1n} \\
0&1&0 &\dots &0& \vdots\\
\vdots&\vdots &\ddots&\ddots&\vdots&\vdots\\
0 &0  &  &\ddots &0 &\vdots \\
0 &0 &0 &\dots &1 &\gamma_{n-1,n}\\
0 &0 &0   &\dots  &0  &\gamma^{-1}_{11}
\endpmatrix
\ ,
\eqno(1.6)
$$
where
$$\eqalignno{
&\gamma_{11}=x_{n1}^{1\over 2}\cr 
&\gamma_{in}= - x_{i1}/x_{n1}
,\ \gamma_{ni}=\gamma_{11}  x_{ni}/x_{n1},\ i=2,\dots,n-1 \cr
&\gamma_{1n}= - {{\gamma_{11}} \over { x_{n1}^2 }} 
\sum^{n-1}_{i=2}x_{ni}x_{i1}\ .\cr}$$
Then a direct computation shows that$$
Ad_\Gamma X=\pmatrix
\varkappa_1 &* &* \\
0&\phi_1(X)&*\\
1 &0  &\varkappa_1
\endpmatrix
\ .
\eqno(1.7)
$$

Thus, the $1$-chop $\phi_1(X)$ can be constructed via the adjoint 
action of the Borel subgroup. This simple observation is crucial
for our generalization of results of [DLNT], [S1], [S2], [EFS] to the
general case of the Toda flows on simple Lie algebras. It suggests
that in the general situation one can use the adjoint 
action of the Borel subgroup and projection on the subalgebra to construct 
an analogue
of the $1$-chop 
$$\phi_1=\pi_{\g'}\circ Ad_\Gamma \eqno(1.8)$$
as a map from a simple Lie algebra $\g$ into its
(semi)simple subalgebra $\g'$ in such a way that this map will respect
the Lie-Poisson and $R$-matrix brackets. Also, as we can see, a 
successive application of such maps to a generic element of
$\epsilon + \b_-$ allows to construct a {\it normal form} similar
to (1.5) and thus to recover the unpublished result by Kostant
on the structure of generic coadjoint orbits.

{\bf 1.3.} Note, that functions $I_{jk}$ defined by (1.2), (1.3) are 
so-called {\it parabolic Casimirs},
i.e. they are $Ad_{\PP_k}$-invariant functions that depend only on the value
of the $k$-chop $\phi_k(X)$ or, in other words, they can be obtained form
the Chevalley invariants of the Levi component of $\PP_k$ .
In the $sl(n)$ case one can oneself to  parabolic Casimirs to construct a
maximal involutive family due to the rather accidental fact that the number
of independent Chevalley invariants in $sl(n)$ is equal to the half of the
difference between dimensions of generic coadjoint orbits in $sl(n)$
and in the image of the $1$-chop, $sl(n-2)$. Roughly speaking, this allows
to disregard ``stars' in (1.7) and to continue constriction of 
Poisson-commuting
integrals using only $\phi_1(X)$. However, it was pointed out in [EFS] that
this approach does not work in $G_2$. This prompted us to use in our 
construction all parabolic invariants rather then only parabolic Casimirs
or, in other words, to use all the data contained in $Ad_\Gamma X$ before
applying $\pi_{\g'}$ in (1.8).

Another phenomenon noticed in [EFS] is the  existence of distinct maximal
involutive families for the generic Toda flows, which can be seen already
in the $sl(4)$ case. This led the authors of [EFS] to conjecture that
the  Toda flows are integrable in the non-commutative sense. Using a  
reduction to of generic elements of $\epsilon + \b_-$ to a normal
form and an interplay between the adjoint and coadjoint actions on
$\epsilon + \b_-$, we shall prove this conjecture (Theorem 5.1).

The paper is organized as follows. In section 2 we explain notations and 
collect necessary facts on special subalgebras of a simple algebra $\g$
that will be used in the sequel. We also provide a background on the 
Toda flows and relevant Poisson structures. In section 3 we define
the $1$-chop map in the form (1.8) and use it to recover Kostant's result 
on a structure of generic coadjoint orbits. In section 4 we prove that
the $1$-chop map is Poisson w.r.t. to both Lie-Poisson and $R$-matrix
brackets. In section 5 we prove the theorem on non-commutative integrability
of the generic Toda flows and also explain that $1$-chop can be viewed
as a Hamilton reduction w.r.t. the Poisson non-commutative family of
parabolic invariants. The latter result make it possible to prove
in section 6 a complete integrability of the Toda flows in $G_2$ . 
In section 7 we use what may be called a nonlinear nongeneric version
of the Manakov-Mischenko-Fomenko shift of argument method to construct
a maximal involutive family of parabolic invariants whose independence
is proved in section 8. Finally, in section 9 we conclude the proof
of our main theorem.

\vskip .2cm

\noindent
{\bf Acknowledgments.} Authors benefited a lot from the stimulating 
discussions
with  A.Bloch, P. Deift, N.Ercolani, H.Flaschka, T.Ratiu
and S.Singer. We are grateful to S. Singer for sharing with us
the notes on Kostant's ``cascade construction''. This work was initiated
while the second author was a visitor at the Department of Mathematics, 
University of Arizona, to which he wishes to express his gratitude 
for financial support.

\head 2. Notations and Preliminaries
\endhead

{\bf 2.1.} Let $\g$ be a simple Lie algebra over $\Bbb C$ of rank $r$ 
different from $sl(r+1)$ , $\G$ is a corresponding Lie group, $\h$ is 
a  Cartan subalgebra of $\g$.
We denote by $\Phi$ the root system of $\g$, by
$\Phi^+$ (resp. $\Phi^-$) the set of all positive (resp. negative)
roots and by $\alpha_1,\dots , \alpha_r$ the basis
of simple positive roots. We also fix a Chevalley basis 
$\{ e_\alpha,\ \alpha\in\Phi; \ h_i,\ i=1,\dots, r\}$ in $\g$ .
All properties of the root systems and Chevalley bases that we will
need can be found in [H]. Recall, in particular, that for
$\alpha, \beta\in \Phi$,  the $\beta$-{\it string}
through $\alpha$ is the the maximal set of roots
of the form $\alpha + i\beta ,\ i\in \Bbb Z$ . Then $i$ ranges
from $-p$ to $q$ and 
$p-q= \langle \alpha | \beta \rangle = 
{{2 (\alpha , \beta)} \over {(\beta,\beta)}}$ .

Let $\b_+ , \n_+,  \b_- ,  \n_-$ be, respectively, the Borel
subalgebra for $h$, its nilradical and their opposites. 
Corresponding groups are denoted by $\B_\pm, \N_\pm$ .
We use notations
$\pi_+=\pi_{\b_+}, \pi_-=\pi_{\n_-},
 \pi_{\b_-},  \pi_{\n_+} $ for the natural projections on $\b_+, \n_-,
 \b_-, \n_+$ .

Following [K1], we define an element
$\epsilon= \sum^r_{i=1} \alpha_i \ $ and realize $\b^*_+$ as an affine
subspace $\epsilon + \b_-$ . Then the coadjoint action of $\B_+$ on
$\b^*_+$ takes the form
$$ Ad^*_b \zeta = \epsilon + \pi_{\b_-} Ad^{-1}_b \zeta \ , $$
where $\  b\in \B_+, \ \zeta \in \epsilon + \b_- $ . 
We denote the coadjoint orbit of $\zeta \in \epsilon + \b_- $ 
by $O_\zeta$ .

{\bf 2.2.} Let $m$ be the maximal positive root and
$$h_m= [e_m, e_{-m}] \ . \eqno(2.1)$$

The following notations will be used throughout the paper:

$$ \eqalignno{
&\Phi^\pm_\F = \{ \alpha \in \Phi^\pm \ :\ ( m, \alpha )  \ne 0\}\cr
&\Phi'=\{ \alpha \in \Phi \ :\ ( m, \alpha ) = 0\} \cr
&\g'= Span \{ e_\alpha : \alpha \in \Phi'\} \cr
&\F_\pm= Span \{ e_\alpha : \alpha \in \Phi^\pm_\F\} &(2.2)\cr
&\P_\pm=\g' + \F_\pm \cr
&\V_\pm= Span \{ e_\alpha : \alpha \in \Phi^\pm_\F, \alpha\ne m\}\cr
&\tilde \F_\pm = Span \{ \F_\pm, h_m\}
\cr }
$$

In the rest of the paper every notation followed by $'$ will have the
same meanilng for $\g'$ as the notation without $'$ has for $\g$ .

It follows from definitions above that
$$ [ e_{\pm m}, \P_\pm] = 0\ , \ [ h_m, \g']=0\ . \eqno(2.3)$$

Clearly, $\Phi'$ is a reduced root system and $\g'$ is a semisimple
subalgebra of $\g$. The Dynkin diagram of $\Phi'$ is obtained by deleting
from the Dynkin diagram of $\Phi$ the vertices that are connected with the
vertex that corresponds to the minimal root $-m$ in the extended 
Dynkin diagram of $\Phi$ . The rank of $\Phi'$ is \ $r-1$\ if 
$\Phi$ is not of type $A_r$ and \ $r-2$\ otherwise (see, e.g. [OV], 
table 6). 

Denote the Killing form on $\g$ by $\langle\ ,\ \rangle$ .
If $\g = \L_1 + \L_2$ is a decomposition of $\g$ into 
the direct sum of linear subspaces, we will identify the dual
space $\L_1^*$ with the subspace $\L_2^\perp$ orthogonal to $\L_1$
w.r.t. the Killing form. In particular,
$$ \F_\pm^* \simeq \F_\mp, \quad \V_\pm^* \simeq \V_\mp, \quad 
\P_\pm^* \simeq \P_\mp \ .$$

If $f$ is a smooth function on $\L_1$ we define 
$\nabla f (\xi) \in \L_2^\perp$ for $\xi \in \L_1$  by 
$$ {{d} \over {d \varepsilon}} f(\xi + \varepsilon\delta \xi)
\restriction_{\varepsilon=0}= \langle \nabla f (\xi), \delta \xi \rangle \ 
.$$

The direct sum decomposition 
$$\g = \F_- + \F_+ + \Bbb C \{ h_m \} + \g' = \tilde \F_- + \P_+=
\P_- + \tilde \F_+  \eqno(2.4)$$
will prove useful in the sequel as will a representation of elements $g\in\g$
in the form
$$ g= x_- e_{-m} + v_- + x_0 h_m + v_+ + x_+ e_m + g'\ , \eqno(2.5)$$
where $v_\pm \in \V_\pm\ , \ g'\in \g' \ $ . We can define
$$\pi_{\V_\pm} g= v_\pm , \  \pi_{\g'} g= g' \ .$$

The properties of $\F_\pm$ can be summarized in the following

\proclaim{Proposition 2.1} 
\roster
\item $\F_\pm$ is a nilpotent subalgebra of dimension
$2 N + 1$ .
\item The following relations hold true
$$ 
[ e_{\pm m},  \V_\mp] = \V _\pm\ ,
\ [\V_\pm, \V_\pm] = \Bbb C \{ e_{\pm m}\ \}, 
 \ [\g', \V_\pm] \subset \V_\pm\ . \eqno(2.6)$$
\item $$[\V_+, \V_-]\cap (\V_- + \V_+) = 0 \ .\eqno(2.7)$$
\item 
$ad_{\g'}\restriction_{\V_\pm}$ is an irreducible representation
of $\g'$ that preserves a nondegenerate skew-symmetric
form $ \langle ad_{e_{_{\mp m}}} \cdot , \cdot \ \rangle  $. All weights
of this representation are nonzero and all weight spaces are
one-dimensional.

\endroster
\endproclaim

\demo{Proof} 
(1) Since $ (m, \alpha) \ge 0$ for any $\alpha \in \Phi^+$ we conclude
that $\F_+$ is a subalgebra. Obviously, it is nilpotent. Moreover, since
for $\alpha \in \Phi^+_\F$\   $ (m, \alpha) > 0$ and 
$m + \alpha \notin \Phi$ , it follows that $m - \alpha$ also belongs to
$\Phi^+_\F$ . Roots $\alpha$ and $m-\alpha$ cannot coincide, therefore
the dimension of $\F_+$ is odd.

(2)
Let $\alpha \in \Phi^+_\F\smallsetminus\{ m \}$ . Then the $m$-string
through $\alpha$ contains only $\alpha, \alpha - m$ . Clearly,
\ $\alpha - m \in\Phi^-_\F\smallsetminus\{- m \}$ \ and the first equality
in (2.6) is proved. Moreover, 
$\langle \alpha | m \rangle = {{2 (\alpha , m)} \over {(m,m)}} = 1$ .
If $ \alpha, \beta, \gamma = \alpha + \beta $ all belong to
$\Phi^+_\F\smallsetminus\{ m \}$ , we have 
$(m,m) = 2 (\gamma,m)= 2 (\alpha + \beta,m)= 2 (m,m)$ which is absurd.
Thus, $\alpha + \beta \in \Phi^+_\F$ if and only if $\alpha + \beta =m$ .
This proves the second equality in (2.6). The third relation follows
from the fact that $ (m, \alpha + \beta)=(m, \alpha)> 0$ for
any $\alpha \in \Phi^+_\F\smallsetminus\{ m \}, \ \beta\in \Phi'$ .

(3) follows from $(m, \alpha - \beta) =  (m,m) -(m,m) =0$ for any
$ \alpha, \beta \in \Phi^+_\F\smallsetminus\{ m \}$ .

(4) Due to the last relation of (2.6), $\g'$ acts on $\V_+$ . 
Since $\g$ is not of type $A_n$, all the roots in $\Phi^+_\F\setminus \{m\}$
has a form $\alpha = \alpha_{i_0} + \sum_{i\ne i_0}n_i \alpha_i$ , where
$\alpha_{i_0}$ is the only simple root connected with $-m$ in the extended
Dynkin diagram of $\g$. It follows that a cyclic subspace of $ad_{\g'}$ 
generated by $e_{\alpha_{i_0}}$ coincides with $\V_+$ . The weight spaces
of $ad_{\g'}\restriction_{\V_+}$ are just the root spaces corresponding to
different $\alpha\in \Phi^+_\F\setminus \{m\}$ . Finally, the statement about
the form $ \langle ad_{e_{_{- m}}} \cdot , \cdot \ \rangle  $ follows from
(2.3) and (2.6).
\qed
\enddemo

\remark{Remark}
In the $sl(n)$ case $\F_+$ is spanned by matrix elements $e_{1i}, e_{jn},\
i=2,\dots, n, j=2,\dots,n-1$ , $\g'$ is $sl(n-2)$ embedded into $sl(n)$
as a central $(n-2)\times (n-2)$ block and $ad_{\g'}\restriction_{\V_+}$
has to irreducible components.
\endremark
\vskip .2cm

Since for any $\alpha \in \Phi^+_\F\smallsetminus\{ m \}$ the 
$\alpha$-string through $m$ is $m, m - \alpha$, we have (see [H],
p. 146) for $v_\pm \in \V_\pm$,
$$ [e_m,[e_{-m}, v_+]]=v_+ \ ,\ [e_{-m},[e_m, v_-]]=v_- \ . \eqno(2.8)$$

{\bf 2.3.} Recall that the Lie-Poisson (Kirillov-Kostant)
bracket $\{\ ,\ \}$ is defined by

$$\{f_1, f_2\} (g)= \langle g, [\nabla f_1 (g), \nabla f_2 (g)]\rangle \ 
\eqno(2.9)$$
and $R$-matrix bracket ([RSTS1], [RSTS1], [STS]) $\{\ ,\ \}_R$ is defined by
$$\eqalignno
{\{f_1, f_2\}_R &(g)=
\langle g,[\pi_+ \nabla f_1 (g), \pi_+ \nabla f_2 (g)]
- [\pi_- \nabla f_1 (g), \pi_- \nabla f_2 (g)]\rangle \cr
&= \langle g, [\nabla f_1 (g), \nabla f_2 (g)] - [\nabla f_1 (g),\pi_- \nabla 
f_2 (g)]-
[\pi_- \nabla f_1 (g), \pi_- \nabla f_2 (g)]\rangle  \ & (2.10) \cr
&= \langle g, - [\nabla f_1 (g), \nabla f_2 (g)] + [\nabla f_1 (g),\pi_+ 
\nabla f_2 (g)]+
[\pi_+ \nabla f_1 (g), \pi_- \nabla f2 (g)]\rangle \ .
\cr }
$$

The Lie-Poisson bracket on $\b_+^*$ is defined by

$$\{f_1, f_2\} (\eta)= \eta ([d f_1 (g), d f_2 (g)]) \ , \eta \in \b_+^*$$
and its pull-back to $\epsilon +\b_-$ can be found as

$$\{\tilde f_1, \tilde f_2\}_{\epsilon +\b_-} (\zeta)=
\langle \zeta , [\pi_+ \nabla f_1 (\zeta), \pi_+ \nabla f_2 (\zeta)] \rangle
\ . \eqno(2.11)$$
Note, that $\left (\epsilon +\b_- , \{\ ,\ \}_{\epsilon +\b_-}\right )$
is a Poisson submanifold of $\left ( \g, \{\ ,\ \}_R \right )$ .
Symplectic leaves of (2.11) are coadjoint orbits of $\B_+$ in $\epsilon +\b_-
$ .

The Hamiltonian $H(\zeta)=\langle \zeta, \zeta \rangle$ generates 
{\it the Kostant-Toda flow} on $\epsilon +\b_-$. The corresponding Hamilton`s
equation has a Lax form

$$\dot \zeta = [ \zeta, \pi_- \zeta] \eqno(2.12)$$
and can be solved via the {\it factorization method}
([K1], [OP], [RSTS1], [Sy]) :

If $\zeta(0)=\zeta_0$ is initial condition for (2.12), consider
the factorization $\exp(t\zeta_0)=n(t)b(t)$ with $n(t)\in \N_-,\ 
b(t)\in \B_+$. Then
$$\zeta(t)= b(t)\zeta_0 b(t)^{-1} \ . \eqno(2.13)$$

The important consequence of (2.13) is that a restriction on $\epsilon +\b_-$
of any $Ad_{\B_+}$-invariant function on $\g$ is an integral of motion
of the full Kostant-Toda flow.

In the proposition below we list well-known properties of brackets
(2.9)--(2.11) that will be used in the next sections. We suggest [RSTS2]
as a reference.

\proclaim{Proposition 2.2} Let $I$ be $Ad_{\G}$-invariant and $f_1,f_2$ be
$Ad_{\B_+}$-invariant functions on $\g$ and $\tilde I,
\tilde f_1,\tilde f_2$ be their restriction on $\epsilon+ \b_-$. Then
\roster
\item 
$$ \{f_1, f_2\}_R=\{f_1, f_2\} \eqno(2.14)$$
\item 
$$\{\tilde f_1 , 
\tilde f_2  \}_{\epsilon + \b_-}=
\{f_1, f_2\}\restriction_{\epsilon + \b_-} \eqno(2.15)$$
\item
$$\{\tilde I , 
\tilde f_1  \}_{\epsilon + \b_-}= 0
\eqno(2.16)$$
\endroster
\endproclaim

\head 3.  Generic Coadjoint Orbits
\endhead

{\bf 3.1.} Let $g\in\g$ is such that in the decomposition (2.5) $x_- \ne 0$.

\proclaim{Proposition 3.1} For any 
nonzero $\lambda \in \Bbb C$,
there exists a unique element $\Gamma \in \exp (\tilde \F_+)$ 
such that 
$$Ad_{_\Gamma} g = \lambda e_{-m} + \phi_1 (g) + F_1(g) := \lambda e_{-m} + P(g)\ ,\eqno(3.1) $$
where $\phi_1 (g)\in \g' ,\ F(g)\in \F_+$ .
\endproclaim

\demo{Proof} Let $g\in\g$ be decomposed as in (2.5).
According to (2.6), (2.7), for any $w \in \V_+$ 
$ ad_w g \in \g - \Bbb C \{ e_{-m} \} , \ 
ad^2_w g \in \g' + \tilde \F_+$ . Therefore, a projection of $\exp(ad_w) g$
on $\V_-$ is equal to $v_- + x_- [w,e_{-m}]$. Put $w=x^{-1}_-[e_m, v_-]$ .
Then, by (2.8)
$$\tilde g = Ad_{\exp(w)} g= \exp(ad_w) g = 
x_- e_{-m}  + \tilde x_0 h_m + \tilde v_+ + \tilde x_+ e_m + \tilde g'\  . $$
Now it is not hard to see that for a suitable $\nu \in \Bbb C$
$$\Gamma =\exp (\nu h_m)  \exp\left (-{{\tilde x_0}\over{x_-}}e_m\right )   
\exp(w)\eqno(3.2)$$
satisfies (3.1).

The uniqueness of $\Gamma$ follows from the fact that no nonzero element
from $\tilde \F_+$ commutes with $ e_{-m}$.
\qed
\enddemo

\definition{Definition 3.1}
We call the map $\phi_1: \g \to \g'$ the {\it $1$-chop}.
\enddefinition

\remark{Remarks} {\bf 1.} In the $sl(r+1)$ case Definition 3.1
coincides with that of [S1,S2]. However, (3.1) in this case should read
$$Ad_{_\Gamma} g = 
\lambda e_{-m} + \phi_1 (g) + F_1(g) + \varkappa_1(g)h_0 \ ,\eqno(3.1') $$
where $h_0$ is the normalized element $\h$ which is orthogonal
to $h_m$ and to the Cartan subalgebra $\h'$ of $\g'$ .

{\bf 2.} One can extend the definition of the $1$-chop to semisimple
algebras by defining $\phi_1$ in the way described above on each 
of the simple components.

{\bf 3.} It follows from a construction of $\Gamma, \phi_1, F_1$, that
``matrix elements'' $\langle F_1 (\cdot ), e_{-\alpha} \rangle , 
\alpha \in\Phi_\F^+ , \  \langle \phi_1 (\cdot ), e_{\gamma} \rangle , 
\ \gamma \in{\Phi'}^+$ considered as functions on $\epsilon + \b_-$
are almost everywhere independent.
\endremark
\vskip .2cm

{\bf 3.2.}The transition from $\Phi$ to $\Phi'$ is the first step of Kostant's
"cascade construction", which can be found in [J], [LW]. If one applies
this step again to each of the irreducible components $\Phi_i'$
of $\Phi'$, one can define maximal roots $m_{1i}\in \Phi_i'$ and 
root systems $(\Phi_i')'$ . Continuation of this procedure until it
eventually ends enables us to construct a maximal set 
$\M = \{ m, m_{1i},...\}$ of {\it strongly orthogonal roots} in $\Phi^+$, 
i.e.
maximal subset of $\Phi^+$ such that for any two roots in it neither their
nor their difference is a root.

Let us enumerate elements of $\M$ \ : 
$\M =\{ \beta_1=m, \dots , \beta_l \}$ .
In view of the last remarks, we can define $1$-chops
$\phi_2: \g' \to \g_2=(\g')',\ \phi_2: \g_2 \to \g_3=(\g_2)',\dots,
\phi_{k+1}: \g_k \to 0$ . Then maximal roots of simple components of 
$\g, \g', \dots, \g_k$ form $\M$. Note that, by construction, each simple
root $\alpha_i$ is not orthogonal to exactly one root in $\M$ and, on the other
hand, each $\beta_j\in \M$ is not orthogonal to either one or two simple roots.
This observation allows one to describe the number $l=Card(\M)$ in 
the following way.

\proclaim{Proposition 3.2} Let $w_0$ be the longest element of the
Weyl group and $s$ be the number of simple roots which are not invariant 
under the action of $-w_0$. Then
$$l + {s\over 2} = r \ .$$
\endproclaim

\demo{Proof} For every $\beta_j\in \M$ , $-w_0 (\beta_j)=\beta_j$ 
( [LW], Lemma 3.1 ) . Then 
$(\beta_j, -w_0(\alpha_i) ) = (\beta_j,\alpha_i)$, which
implies that  $-w_0$ interchanges $\alpha_i,\alpha_{i'}$ which
are not orthogonal to the same $\beta_j\in \M$ and fixes 
all other simple roots.  The statement follows.
\qed
\enddemo

Denote
$$\eqalignno{
&\phi^{(j)} = \phi_j\circ\phi_{j-1}\circ\cdots\circ\phi_1\ ,\ j=1,\dots,k\cr
&\phi^{(k+1)} = F_{k+1}\circ\phi^{(k)}\ ,\cr}
$$
where $F_j,\ j=1,\dots,k+1$ can be defined similarly to (3.1).

\definition{Definition 3.2}
We call an element $g\in\g$ {\it generic} if maps
$\phi^{(j)}, \ j=1,\dots,k+1$ are well defined on $g$ .
A coadjoint orbit $O_\zeta$ in $\epsilon + \b_-$ is called
generic if $\zeta$ is generic.
\enddefinition

Let now $\zeta \in \epsilon +\b_-$ be generic and let
$\epsilon' = \sum_{(\alpha_i, m)=0} e_{\alpha_i}$ .
 Then $ \zeta' =
\epsilon' +  \pi_{\b_-} \phi_1 (\zeta)$ is a generic element
in $\g'$.

\proclaim{Proposition 3.3} 

$\dim O_\zeta - \dim O_{\zeta'} = \dim \tilde \F_+ = 2 N + 2 \ $

$ \operatorname{codim} O_\zeta = \operatorname{codim} O_{\zeta'}$ . 
\endproclaim

\demo{Proof} By Proposition 3.1 there exists unique
$\Gamma \in \exp (\tilde \F_+)$ such that $\zeta_\lambda =
Ad^*_{\Gamma^{-1}} \zeta = 
\lambda e_{-m} + \pi_{\b_-} \phi_1 (\zeta) + \epsilon$ .
Then \ (i) for any $v_-\in V_-; x_-\ne 0, x_0\in\Bbb C$ there exists 
$\xi\in O_\zeta$ such that $\pi_{\tilde \F_+} \xi = 
x_-e_{-m} + v_- + x_0 h_m$ ; \  
(ii)\ $(Ad^*_b \zeta_\lambda - \zeta_\lambda) \in \g'$ implies
$b\in \B'_+=\B_+ \cap \G'$ . 

Therefore, $ O_{\zeta'}\simeq Ad^*_{\B'_+} \zeta_\lambda$
and
$O_\zeta =Ad^*_{\exp (\tilde \F_+)} Ad^*_{\B'_+} \zeta_\lambda
\simeq O_{\zeta'}\times\Bbb C^{2 N +1}\times\Bbb C^*$ . 
\qed
\enddemo

\remark{Remark} If  $\g$ is $sl(r+1)$ then
$ \operatorname{codim} O_\zeta - \operatorname{codim} O_{\zeta'} = 1$, since
$\varkappa_1$ in (3.1$'$) is a coadjoint invariant.
\endremark

In the table below we have listed for all simple algebras the type of $\g'$
and the difference $\Delta\dim$ of dimensions of generic coadjoint
orbits $O_\zeta$  and $O_{\zeta'}$ .

$$\matrix
\g  \hfill       &A_n \hfill   & B_n  \hfill       & C_n \hfill   & D_n 
\hfill       &E_6&E_7&E_8&F_4&G_2 \\
\g' \hfill       &A_{n-2}& A_1+B_{n-2} & C_{n-1}& A_1+D_{n-
2}&A_5&D_6&E_7&C_3&A_1 \\
\Delta\dim &2n \hfill    & 4n-4 \hfill       & 2n  \hfill   & 4n-6 \hfill      
&22 &34 &58 &16 &6
\endmatrix
$$

Repeatedly applying Propositions 3.1, 3.3  to $\g',\g_2,\dots, \g_k$ 
we recover the 
following result due to Kostant :

\proclaim{Theorem 3.1} Let $\h_0$ be an orthogonal complement in $\h$
to $Span \{ [e_{-\beta}, e_{\beta}] , \beta \in \M \}$ .
If $\zeta$ is generic then there exist the unique element
$h_0 \in \h_0$ such that 
$$\zeta_0=\sum_{i=1}^l e_{-\beta_i} + h_0 +\epsilon \in O_\zeta\ .$$
The codimension of $O_\zeta$ in $\epsilon +\b_-$ is equal to the dimension
of the stabilizer of $\zeta_0$ and is $r-l$ .
\endproclaim

\remark{Remark} Invoking Proposition 3.2 we obtain  that the 
codimension of the generic orbit is $s/2$ . This result was obtained by
Trofimov in [Tr2] .
\endremark
\vskip .2cm

Due to Theorem 3.1 for any generic $\zeta\in \epsilon + \b_-$ 
there exist 
$b_\zeta\in\B_+$ such that 
$$Ad_{b_\zeta}^* \xi =
\zeta_0=\sum_{i=1}^l e_{-\beta_i} + h_0+\epsilon\ , \eqno(3.3)$$
where $h_0\in\h_0$ .
Denote $\h_1=\h_0^\perp=Span \{ [e_{-\beta}, e_{\beta}] , \beta \in \M \}$ 
and consider the factorization $\T=\T_0 \T_1$
of the maximal torus $\T$ corresponding to the linear space decomposition
$\h = \h_0 + \h_1$ .
Then a stabilizer of $\zeta_0$ under the coadjoint action is $\T_0$
and $b_\zeta$ is defined uniquely up to a right multiplication by
elements of $\T_0$ . Let $b_\zeta=\tilde b_\zeta t_\zeta$ be a 
factorization of $b_\zeta$ into
the product of unipotent $\tilde b_\zeta$ and $t_\zeta\in \T$. 
We can make the choice of $b_\zeta$ unique
by demanding that  $t_\zeta$ belongs to $\T_1$ and thus establish
a one-to-one correspondence between $\N_+  \T_1$ and $O_\zeta$ :

$$ \N_+  \T_1 \ni \tilde b\ t_1 \leftrightarrow 
Ad^*_{(\tilde b\ t_1)^{-1}} \zeta_0 \in O_\zeta\ .\eqno(3.4)$$

We finish this section with the following technical lemma.

\proclaim{Lemma 3.1} For $\nu \in \n_-$, $[\nu, \zeta_0] \in\h_0$
implies $\nu=0$ .
\endproclaim
\demo{Proof}
Let us represent $\nu$ as a sum $\nu=\nu_{-1} + \nu_{-2} + \cdots$ , where
$\nu_{-i}$ belongs to the subspace $V_{-i}$ of $\g$ generated 
by roots of height $-i$ .
Then 
$$ \eqalignno{
&[\nu_{-1},  \epsilon ] \in \h_0 \cr
&[\nu_{-2},  \epsilon ]= - [\nu_{-1}, h_0]\cr
&\cdots & (3.5)\cr
&[\nu_{-i-1},  \epsilon ]= l_i (\nu_{-1},\dots, \nu_{-i})\cr
&\cdots \cr }
$$
Here 
$$l_i (\nu_{-1},\dots, \nu_{-i}) = - [\nu_{-i}, h_0] - 
\sum_{\operatorname {height}(\beta_j) + k =i} [\nu_{-k}, e_{-\beta_j}]\ .$$
It is shown in [K2,K3] that $ad_\epsilon : V_{-i-1}\to V_{-i}$ is injective. 
Therefore, (3.5) implies that $\nu_{-2}, \nu_{-3}, \dots$ are uniquely
determined by $\nu_{-1}$ and , therefore, the number of linearly independent
solutions to (3.5) cannot exceed the number of linearly independent
$\nu_{-1}$ satisfying the first relation of (3.5), the latter being equal 
to $\dim \h_0$ . Note that for all root systems but $A_n, D_{2n+1}$ and
$E_6$ ,  $\dim \h_0=0$ . The remaining three cases can be checked by a direct
computation.
\qed
\enddemo

\head 4. $1$-Chop Is Poisson
\endhead

{\bf 4.1.} Our next objective is to study a behaviour of the
1-chop map $\phi_1(g)$ and , more generally,
$P(g)$ with respect to Poisson brackets (2.9), (2.10). 

First, we compute a variation $\delta P$ \ :
$$ \delta P = Ad_{_\Gamma}\delta g  - [\lambda e_{-m} + P, \delta \gamma]\ 
,$$
where $ \delta \gamma = R_{_{\Gamma^{-1}}*}\delta \Gamma$ .
Since 
$$\delta P \perp \tilde \F_+\ ,\eqno(4.1)$$ 
we have for any $\psi \in \tilde\F_+$

$$\langle Ad_{_\Gamma}\delta g , \psi \rangle  
= \langle [\lambda e_{-m} + P,\delta \gamma] , \psi\rangle  
=\lambda \langle [e_{-m},\delta \gamma] , \psi\rangle  \ . \eqno(4.2)$$

Let $\delta \gamma = v_\gamma + \gamma_+ e_m + \gamma_0 h_m$ . Then , by 
(2.8), for any $v\in \V_+$,
$$\langle Ad_{_\Gamma}\delta g , v \rangle  = \langle [Ad_{_\Gamma}\delta g , 
e_m ], [e_{-m}, v] \rangle  =
- \lambda \langle v_\gamma, [e_{-m}, v]\rangle  \ .$$
Therefore,
$$v_\gamma = \lambda^{-1} \pi_{\V_+} [e_m , Ad_{_\Gamma}\delta g ] \ . 
\eqno(4.3)$$

To find $\gamma_0$ , substitute $\psi=e_m$ into (4.2) :
$$\langle Ad_{_\Gamma}\delta g , e_m \rangle  
= \lambda \gamma_0 \langle [e_{-m}, h_m], e_m\rangle = 
\lambda \gamma_0 \langle h_m, h_m\rangle  \ .$$
Thus,
$$ \gamma_0 = \lambda^{-1} {{\langle Ad_{_\Gamma}\delta g , e_m \rangle } 
\over {\langle h_m, h_m\rangle }} \ .\eqno(4.4)$$
Using (2.6), (4.1), we obtain

$$\delta P = \pi_{\P_+} Ad_{_\Gamma}\delta g - [P, v_\gamma +\gamma_0 h_m] \ 
,$$
where $v_\gamma$ and $\gamma_0$ are given by (4.3), (4.4).

Now we can compute $\nabla (f\circ P) (g)$ for a function $f \in C^\infty 
(\P_+)$\ .
Indeed,
$$\eqalignno
{&\delta (f\circ P)= \langle \nabla f (P), \delta P\rangle \cr
&= \langle \delta f (P), \pi_{\P_+} Ad_{_\Gamma}\delta g - 
\lambda^{-1} [P, \pi_{\V_+} [e_m , Ad_{_\Gamma}\delta g ] 
- \gamma_0 [P, h_m ]\rangle  \cr
&= \langle \nabla f(P), Ad_{_\Gamma}\delta g \rangle  - 
\lambda^{-1} \langle \pi_{\V_-} [\nabla f (P), P], Ad_{_\Gamma}\delta 
g\rangle   -
\lambda^{-1} {{\langle Ad_{_\Gamma}\delta g , e_m \rangle } \over {\langle 
h_m, h_m\rangle }} \langle [\nabla f (P), P], h_m\rangle  \cr
&= \langle Ad^{-1}_{_\Gamma} \left ( \nabla f(P) + \lambda^{-1} [e_m , 
\pi_{\V_-} [\nabla f (P), P]]
- \lambda^{-1} {{\langle [\nabla f (P), P], h_m\rangle } \over {\langle h_m, 
h_m\rangle }} e_m \right ), \delta g \rangle  \ .\cr}
$$

Therefore,
$$ \nabla (f\circ P) (g) = 
Ad^{-1}_{_\Gamma} \left ( \nabla f(P) + \lambda^{-1} (v_f + x_f  e_m) \right 
)\ , \eqno(4.5)$$
where \ 
$$v_f = [e_m , \pi_{\V_-} [\nabla f (P), P]] \in \V_+\ \operatorname{and} \
\ x_f={{\langle [\nabla f (P), P], h_m\rangle } \over {\langle h_m, 
h_m\rangle }}\ .$$

Let us consider a particular case  when a function $f$ belongs to 
$C^\infty (\g')\subset C^\infty (\P_+)$. Then $\nabla f \in \g'$\  and\ 
$(f\circ P) (g) = (f\circ \phi_1) (g)$. Therefore,  by (2.6), (3.6), 
\newline
$v_f = 0$, \ $x_f=0$ , \ 
$\nabla (f\circ \phi_1) (g) \in \P_+$ ,\ 
$\pi_{\g'} \nabla (f\circ \phi_1) (g) = \nabla f(\phi_1)$ 
and
$$\pi_- \nabla (f\circ \phi_1) (g) = \pi_-\nabla f(\phi_1)= 
Ad^{-1}_{_\Gamma}\pi_-\nabla f(\phi_1) \ .\eqno(4.6)$$

These observations enable us to prove the following

\proclaim{Theorem 4.1} The map $\phi_1: \g \to \g'$, defined by (3.1) is 
Poisson
w.r.t.  both Lie-Poisson and $R$-matrix brackets.
\endproclaim

\demo {Proof} Let $f_1, f_2 \in C^\infty (\g')$ . Then
$$\eqalignno
{\{ f_1\circ \phi_1 , f_2\circ \phi_1 \} (g)&= \langle Ad_{_\Gamma} g,
[\nabla f_1(\phi_1) + \lambda^{-1}x_{f_1} e_m,
\nabla f_2(\phi_1) + \lambda^{-1}x_{f_2} e_m]\rangle \cr
&= \langle \phi_1, [\nabla f_1(\phi_1), \nabla f_2(\phi_1)]\rangle \ . \cr} 
$$
The second part of the statement can be proved similarly, if one uses (4.6) 
and the second line
of (2.10).
\qed
\enddemo

{\bf 4.2.} Next, we compute a push-forward of the Lie-Poisson
bracket (2.9) on $\g$ under the map $P$ .
For any $f_1, f_2 \in C^\infty (\P_+)$ ,  (3.6) implies

$$ \eqalignno
{\{ f_1\circ P , f_2\circ P \} (g) &= 
\langle Ad_{_\Gamma} g, [\nabla f_1(P) + \lambda^{-1} (v_{f_1} + x_{f_2}  
e_m),
\nabla f_2(p) + \lambda^{-1} (v_{f_2} + x_{f_2}  e_m)] \rangle \cr
&= \lambda^{-1} \langle  e_{-m}, [ v_{f_1} + x_{f_1}  e_m, v_{f_2} + x_{f_2}  
e_m] \rangle 
+\langle P, [\nabla f_1(P), \nabla f_2(P)] \rangle  \cr
&\quad +\lambda^{-1}\langle p, [\nabla f_1(p),v_{f_2} + x_{f_2}  e_m] + 
[v_{f_1} + x_{f_1}  e_m, \nabla f_2(P)]\rangle 
\cr
&= I + II + III \ .\cr}
$$
Here we used the fact that $e_{-m} \perp \P_-$ . Due to (2.6), (2.8)
$$ \lambda I=  \langle  e_{-m}, [ v_{f_1} , v_{f_2} ] \rangle  $$
and
$$ \lambda III = \langle  P, [\nabla f_1(P),v_{f_2}] + [v_{f_1}, \nabla 
f_2(P)]\rangle  \ .$$

Moreover,
$$\eqalignno
{\langle  e_{-m}, [ v_{f_1} , v_{f_2} ] \rangle  &= 
\langle [e_{-m}, [e_m , \pi_{\V_-} [\nabla f_1 (P), P]] ], v_{f_2}\rangle  
\cr
&= \langle\pi_{\V_-} [\nabla f_1 (P), P]] , v_{f_2}\rangle  \cr
&= - \langle [ P, \nabla f_1 (P)], v_{f_2}\rangle \ .\cr}
$$

Thus,
$$\eqalignno
{\lambda (I + III) &= \langle  P,[v_{f_1}, \nabla f_2(P)]\rangle  \cr
&= \langle[e_m , \pi_{\V_-} [\nabla f_1 (P), P]] , [\nabla f_2 (P), P] 
\rangle \cr
&= \langle[e_m , \pi_{\V_-} [\nabla f_1 (P), P]] , \pi_{\V_-} [\nabla f_2 
(P), P] \rangle  \ . \cr}
$$

Denote
$$\eqalignno{
& \{ f_1 , f_2\}_0 (p) = \langle p, [\nabla f_1(p), \nabla f_2(p)] \rangle  
\ , \cr
&\{ f_1 , f_2\}_1 (p) = 
\langle ad_{e_m} \pi_{\V_-} [\nabla f_1 (p), p] , \pi_{\V_-} [\nabla f_2 (p), 
p] \rangle  \ .&(4.7)\cr}$$

We proved

\proclaim{Proposition 4.1} For any $f_1, f_2 \in C^\infty (\P_+)$
$$ \{ f_1\circ P , f_2\circ P \} = \{ f_1 , f_2\}_0 \circ P + 
\lambda^{-1} \{ f_1 , f_2\}_1 \circ P \ := \{ f_1 , f_2\}^\lambda\circ P\  .
\eqno(4.8)$$
\endproclaim

Since $\lambda$ is arbitrary and $P : \g \to \P_+$ is clearly surjective, we 
obtain 
as an immediate corollary

\proclaim{Proposition 4.2} $\{\  ,\ \}_0$ and $\{\  ,\ \}_1$ are compatible 
Poisson brackets.
\endproclaim

\head 5. Non-Commutative Integrability of Toda Flows
\endhead

{\bf 5.1.} The notion of non-commutative integrability we are referring to
in this section goes back to results of Nehoroshev [N] . It is more
general then 
{\it linear} non-commutative integrability introduced by 
Mischenko and Fomenko [MiFo1]. For examples  and bibliography, we refer to
chap. 5 of the book [Fo]. 

Let $\A$ be a  Poisson subalgebra of first integrals
of a Hamilton flow on $2n$-dimensional symplectic manifold.

\definition{Definition 5.1}
A flow is called integrable in
the non-commutative sense 
if $\A$ is generated by functionally independent
functions $f_1,\dots, f_{2n-m}$ such that $\{f_i,f_j\}=0$ for any
$i=1,\dots,m ;\ j=1,\dots,2n-m $ . 

\enddefinition

Consider a generic coadjoint orbit $O_\zeta$ through 
$\zeta\in \epsilon + \b_-$ .

\proclaim{Theorem 5.1} The Poisson subalgebra $\A$ of first integrals
of the Toda flow on $O_\zeta$ is generated by restrictions
to  $O_\zeta$ of $Ad_{\B_+}$-invariant functions on $\g$ and has a
functional dimension $\left (\dim O_\zeta - r\right )$ . Its center is generated by 
restrictions of the Chevalley invariants of $\g$ . The Toda flow
is, therefore, integrable in the non-commutative sense.
\endproclaim

\demo{Proof}
For a generic element $g\in\g$ , we define $b_g$ to be equal to $b_\zeta$, 
where $\zeta= \pi_{b_-} (g) + \epsilon \in \epsilon + \b_-$ and $b_\zeta$
is the unique element of $\N_+  \T_1$ satisfying (3.3) .

Note, that if $b_1$ is an element from $\B_+$ and $b_1=\tilde b_1\ t_0 t_1$ is
its factorization into the  product of unipotent $\tilde b_1$ , 
$t_0\in\T_0$ and $t_1\in\T_1$, then
$$b_{_{(Ad_{b} g)}} 
= (\tilde b t_0 t_1 \tilde b_\zeta t^{-1}_0 t^{-1}_1)\ t_1 t_\zeta =
b b_g t^{-1}_0 \ . \eqno(5.1)$$

For any   $\alpha\in\Phi^+$, define a function
$$\varphi_\alpha (g) = 
\langle  g, Ad_{b_g}e_{-\alpha} \rangle \ . \eqno(5.2)$$
It follows from (5.1) that $\varphi_\alpha (g)$ is semiinvariant under
the adjoint action of $\B_+$ :
$$\varphi_\alpha (Ad_b g) 
= \langle  g, Ad_{b_g}Ad^{-1}_{t_0} e_{-\alpha} \rangle
= \chi_\alpha(t_0)\ \varphi_\alpha (g)\ .$$
Moreover, if $\bold k$ is an integral vector such that 
$\nu=\sum_{\alpha\in\Phi^+} k_\alpha \alpha,
\ k_\alpha\in \Bbb Z$ 
annihilates $\h_0$ then a function 
$$\theta^{\bold k}(g)= 
\Pi_{\alpha\in\Phi^+} \left (\varphi_{\alpha}(g)\right )^{k_\alpha} 
\eqno(5.3)$$
is $Ad_{\B_+}$-invariant on $\g$ .

The number $K$ of linearly independent vectors $\bold k$
such that $\nu$ annihilates $\h_0$ is equal to the number of positive roots
minus the rank of the matrix 
$(\alpha(\eta_i))_{\alpha\in\Phi^+, i=1,\dots,\dim \h_0}$, where 
$\eta_i,\ i=1,\dots,\dim \h_0$ is a basis of $\h_0$. Clearly, this rank is
equal to $\dim \h_0=r-l$ . Thus
$$K=Card (\Phi^+) + l - r = \dim O_\xi - r\ .$$

Now we have to prove that functions $\theta^{\bold k}$ with 
linearly independent vectors $\bold k$ are independent on a generic 
coadjoint orbit $O_\zeta$ . Since such $\theta^{\bold k}$ are obviously
independent as polynomials in $\varphi_\alpha,\ \alpha\in \Phi^+$, 
it suffices to show that $\varphi_\alpha$ are independent on $O_\zeta$ .
To this end, we use
the identification (3.4) and estimate the dimension of the linear space
generated by differentials at identity of $\varphi_\alpha$ considered as
functions on $\N_+ \T_1$ . Differentiating 
$$\varphi_\alpha (Ad^*_{\left (\exp(\varepsilon \delta b)\right )^{-1}} 
\zeta_0 )= 
\left \langle  \pi_{\b_-} (Ad_{\exp(\varepsilon \delta b)} \zeta_0)+ 
\epsilon , Ad_{\exp(\varepsilon \delta b)} e_{-\alpha} \right\rangle \ \ 
,\ \delta b \in \n_+ + \h_1\  $$ 
with respect to $\varepsilon$ at $\varepsilon=0$ , we obtain
$$ \delta \varphi_\alpha = 
\langle \zeta_0, [\delta b , e_{-\alpha}] \rangle=
\langle \delta b , [  e_{-\alpha}, \zeta_0 ] \rangle
\ .$$
Thus, linear dependency of $d \varphi_\alpha$ at identity is equivalent
to existence of nonzero $\nu\in \n_-$ such that 
$ [\nu, \zeta_0 ]\in \h_0 $ which would contradict Lemma 3.1. Thus, the
functional dimension of $\A$ is greater or equal then $\left ( \dim O_\zeta
- r\right )$ .

To complete the proof, one should notice that
if $I_i, \ i=1,\dots,r$ are independent  invariant polynomials on $\g$, then
$I_i(\zeta)= I_i(Ad^{-1}_{b_\zeta} \zeta)$ can be expressed via functions 
$\theta^{\bold k}$ and are in involution with any of them 
(cf. Proposition 2.2). Finally, if there exists either one more integral 
independent of  constructed above or one more element of the center of
$\A$ independent of $I_i, \ i=1,\dots,r$, this is in contradiction with
existence of $r$ independent Hamiltonian flows generated by 
$I_i, \ i=1,\dots,r$.
\qed
\enddemo

\remark{Remark} A non-commutative integrability makes possible existence
of distinct maximal {\it commutative} families for the Toda flows
(phenomenon observed in [EFS]) . Another important implication is that,
due to Nehoroshev's theorem [N], the Toda flow is {\it degenerate}, i.e. its 
trajectories  lie on invariant
manifolds of dimension $r< 1/2 \dim O_\zeta$ . In particular, this enables
one to reconcile Hamiltonian and gradient structures of the full Toda flows.
This aspect is explored in the forthcoming paper [BG].

\endremark
\vskip .2cm

{\bf 5.2.} Theorem 5.1 allows, in principle, to construct
a maximal involutive family of {\it smooth} almost everywhere
independent first integrals of the Toda flow
(cf. [N], [Fo], chap. 5). However, if we are looking for a maximal
family of {\it rational} integrals, it may be helpful to restrict ourselves
to a subalgebra of $\A$ by imposing additional symmetry conditions.

Let $\PP$ be the  parabolic subgroup that stabilizes the weight $m$
(the corresponding subalgebra is $\P_+ + \Bbb C \{ h_m \}$ ) .
It is easy to see that if $f_1$ is an $Ad_{\PP}$-invariant function on $\g$
and $f_2$ is such that  $\nabla f_2(g)\in \P_+ + \Bbb C \{ h_m \}$, then
$ \{ f_1, f_2 \} = 0$ . In particular, we have 

\proclaim{Lemma 5.1} If $f_1$ is an $Ad_{\PP}$-invariant function on $\g$
and $f_2\in C^{\infty}(\g')$, then
$$ \{ f_1, f_2\circ \phi_1 \} = 0\  . \eqno(5.4) $$ 
\endproclaim

Recall that, due to Proposition 2.1, the adjoint action of $\G'$ preserves
$\P_+$.

\proclaim{Lemma 5.2} If $f$ is an $Ad_{\G'}$-invariant function on $\P_+$,
then $f\circ P$ is  an $Ad_{\PP}$-invariant function on $\g$ .
\endproclaim

\demo{Proof} Let $g,\Gamma$ be as in Proposition 3.1 and let $\bold p$ belong
to $\PP$ . Factor $\bold p \Gamma^{-1}$ into the product
$\bold p \ \Gamma^{-1} = \Gamma_1^{-1} \bold g'$ , where 
$\Gamma_1\in \exp (\tilde \F_+)$ and $\bold g'\in \G'$ .
Then $Ad_{\bold p} g= Ad_{\Gamma_1^{-1} \bold g'} (\lambda e_{-m} + P(g))=
Ad_{\Gamma_1^{-1}} (\lambda e_{-m} + Ad_{\bold g'}  P(g)) $ . Then it follows
from Proposition 3.1 that $P(Ad_{\bold p} g)=Ad_{\bold g'}  P(g)$ and, 
therefore, $f(P(Ad_{\bold p} g))=f(P(g))$ .
\qed
\enddemo

Consider now algebra  $\A'$ of $Ad_{\B'_+}$-invariant functions on $\g'$ 
constructed in the same way as $\A$  on $\g$. Then, by Theorem 4.1
$\A'\circ\phi_1$ is a Poisson subalgebra of $\A$ . The same is true
for $\A_{\P_+}\circ P$ , where $A_{\P_+}$ is an algebra of 
$Ad_{\G'}$-invariant functions on $\P_+$ . Moreover, it follows from
Lemmas 5.1, 5.2 that 
$$ \{ f_1\circ P, f_2\circ \phi_1 \} = 0 \eqno(5.5)$$
for any $f_1\in \A_{\P_+}, f_2\in \A'$ .

Let $\A_1$ be a Poisson subalgebra of $\A$ generated by 
$\A'\circ\phi_1 , \A_{\P_+}\circ P$ . Then  Theorem 4.1 and (5.5)
imply that functions $I'_j\circ\phi_1, \ j=1,\dots,r-1$ , 
where $I'_j$ are the Chevalley invariants of $\g'$ lie in the center of $\A_1$ 
along with the Chevalley invariants $I_i, \ i=1,\dots,r$ of $\g$ . This allow
us to restrict all the considerations below to the case when both $g$ and
$\phi_1(g)$ are {\it regular} elements of $\g$ and $\g'$ . In particular,
functions from $\A_{\P_+}$ we are going to consider will be defined on
$\P^r_+=\{ p=g'+ v_+ + x_+ e_m : g'\in \g'\ \operatorname{is \ regular}; \ 
v_+ \in \V_+ \}$ .

Note that a linear function
$$l_m (p) = x_+ ={{\langle e_{-m},p\rangle } \over
{\langle e_{-m},e_m\rangle} }\   \  \eqno(5.6)$$
is $Ad_{\G'}$-invariant due to (2.3).


To compute the functional dimension of $\A_1$ we need

\proclaim{Lemma 5.3} There exist $\left (\dim \F_+ - r \right )$
functionally independent over $\V_+$\  $Ad_{\G'}$-invariant 
functions on $\P^r_+$ .
\endproclaim
\demo{Proof}  
Write $p\in \P^r_+$ as $p=g'+ v + x_+ e_m$ , where $g'=Ad_{\bold g'} h'$ 
and $h'$ is a regular element of  the Cartan subalgebra 
$\h'$ of $\g'$ . Then  $\bold g'$ is defined uniquely up to the right
multiplication by an element of the maximal torus $\T'$ of $\g'$ . 

Let $\bold k=(k_\alpha)_{\alpha\in\Phi_\F^+}$ be a nonnegative 
integral solution of the equation
$$\sum_{\alpha\in\left(\Phi_\F^+ \setminus\{m\}\right )} 
k_\alpha \alpha = k_m m \ . \eqno(5.7)$$
Then, since $m(\h')=0$, the monomial
$$\tilde f_{\bold k} (v) = 
\Pi_{\alpha\in\left(\Phi_\F^+ \setminus\{m\}\right )} 
\langle v, e_{-\alpha} \rangle \ \eqno(5.8)$$
is an $Ad_{\T'}$-invariant function on $\V_+$ and, therefore
function
$$f_{\bold k} (p) = \tilde f_{\bold k} (Ad^{-1}_{\bold g'} v) \eqno(5.9)$$
is $Ad_{\G'}$-invariant. The number of independent
monomials $\tilde f_{\bold k}$ is equal to the number of linearly independent
solutions of (5.7), which , in turn, is equal to 
$\left (\dim \F_+ - r \right )$ . 
\qed
\enddemo

It should be emphasized that, due to the Remark 3 in the end of section 3.1,
functions $f_{\bold k}\circ P, l_m\circ P$ are independent of functions
from $\A'$. Therefore, Theorem 5.1, Lemma 5.3 and property (5.5) lead us
to the following

\proclaim{Proposition 5.1} The functional dimension of the 
Poisson subalgebra  $\A_1$ is equal to
$\left (\dim O_\zeta - 2r+1\right )$ . The center of $\A_1$ is
generated by $2r-1$ functions $I_i, \ i=1,\dots,r$ and 
$I'_j\circ\phi_1, \ j=1,\dots,r-1$ .

If $\C'$ (resp. $\C_{\P_+}$) is a maximal involutive subalgebra
of $\A'$ (resp. $\A_{\P_+}$) then subalgebra $\C$ generated by
$\C', \C_{\P_+}$ is the maximal involutive subalgebra of $\A_1$
whose functional dimension is equal to $1/2 \dim O_\zeta$ .
\endproclaim

Proposition 5.1 suggests an inductive procedure of constructing the 
maximal Poisson commutative family of integrals for the Toda flows in $\g$. 
Indeed, 
if we know how to construct $\C_{\P_+}$ then we are left with the problem
of constructing  the maximal Poisson commutative subalgebra in $\A'$, which
is the same as our initial problem but for the algebra $\g'$ of smaller
rank. 

We shall construct $\C_{\P_+}$ in sections 7, 8. But before  we would
like to show that results of this section already enable us to prove
the complete integrability of the Toda flows in the particular case
when $\g$ is of type $G_2$ .

\head 6. Example: Generic Toda Flows in $G_2$ Are Completely
Integrable.
\endhead

Let $\alpha_1$ be the short and $\alpha_2$ be the long simple positive
roots of $G_2$ . Then $m=2 \alpha_2 +  3 \alpha_1$ and
$\Phi^+_\F= \{ m, \alpha_2 + i \alpha_1\ (i=0,1,2,3) \}$ . $\g'$ is
an algebra of type $A_1$ generated by $e_{\pm\alpha_1}, h_1$ and the maximal
family of strongly orthogonal roots consists of $m$ and $\alpha_1$ . 

According to Theorem 3.1 the dimension of a generic coadjoint orbit in 
$\epsilon +\b_-$ is equal to $8$. Therefore, we need $4$ independent integrals
in involution. We can choose $3$ of them to be the Chevalley invariants 
$I_1, I_2$ of $G_2$ and a superposition $I'_1\circ \phi_1$ of the Chevalley
invariant $I'_1$ of $A_1$ with the $1$-chop map $\phi_1$ .

Let us now choose generators of the algebra $A_{\P_+}$ of 
$Ad_{\G'}$-invariant functions on $\P_+$ . Denote 
$$v_i (v) = \langle e_{\alpha_2 + i \alpha_1}, v \rangle \ 
(i=0,1,2,3)\ . $$
Then functions 
$$\tilde f_1(v)= v_0 v_3, \ \tilde f_2(v)= v_1 v_2, 
\ \tilde f_3(v)= v_0 v_2^3\eqno(6.1)$$
form a maximal family of independent $\h'$-invariant functions on $\V_+$ .
By  (5.6), (5.9), $A_{\P_+}$ is generated by 
$f_i(p)=\tilde f_i(Ad_{\bold g'}v),\
i=1,2,3$ and functions $l_m(p)$ and $I'_1(g')$ . Here $g', \bold g'$ and
$v$ are defined as in the proof of Lemma 5.3.

Due to (5.5), $I'_1\circ \phi_1$ is involution with any function from
$A_{\P_+}\circ P$ . The same is true for $I_1, I_2$ . However, the functional
dimension of $A_{\P_+}\circ P$ is $5$ and, therefore, any function $J$ from
$A_{\P_+}\circ P$ which is independent with $I_1, I_2, I'_1\circ \phi_1$
can be used as the fourth integral needed for complete integrability and,
moreover, if we pick another function, $\tilde J$, as the fourth integral,
the Poisson bracket of $J$ and $\tilde J$ does not have to be zero.
In particular, $J$ can be chosen to be either one of the functions
$f_i\circ P$ obtained from (6.1).

Finally, we present another choice of the fourth integral $J$ that does not
involve the adjoint action of $\bold g'$. Recall (Proposition 2.1) that 
the adjoint action of $\G'$ on $\V_+$ preserves a nondegenerate skewsymmetric
form $ \langle ad_{e_{_{m}}} \cdot , \cdot \ \rangle $ that endows $\V_+$
with a constant  symplectic structure. The action of $\G'$ is Hamiltonian.
For an element $g'\in\g'$ the  flow of $Ad_{\exp(g't}$ is a  Hamiltonian
flow generated by 
$H_{g'}(v)= {1\over2}\langle ad_{e_{-m}}  ad_{g'} v, v \rangle$ . Then 
the moment map $\Cal J : \V_+ \to \g'$ is given by
$$\Cal J (v) = {1\over2} \pi_{\g'} \left ( [v, ad_{e_{-m}} v] \right )
=  {1\over2}\left ( [v, ad_{e_{-m}} v] \right ) \ .$$
(The last equality can be derived from (2.3).)

It follows that function
$$f(p)=f(v)= \langle [ad_{e_{-m}}v,v], [ad_{e_{-m}}v,v] \rangle \ .
\eqno(6.2)$$
is $Ad_{\G'}$-invariant. In the case of $G_2$, a direct computation
shows that function $J=f\circ P$ is nontrivial and  idependent with 
$I_1, I_2, I'_1\circ \phi_1$ . Invoking (5.5), we conclude that functions
$I_1, I_2, I'_1\circ \phi_1,J$ form a maximal family of independent integrals
in involution.

\head 7. Involutive Integrals
\endhead

Recall that in Proposition 4.1 we defined a Poisson bracket 
$\{ \ ,\ \}^\lambda$ on $\P_+$ as a push-forward of the Lie-Poisson
bracket on $\g$ under the map $P$.
In this section we construct a family of $Ad_{\G'}$-invariant and Poisson
commutative with respect to $\{ \ ,\ \}^\lambda$ functions
on $\P_+$ . Then, by Proposition 4.1 and Lemma 5.2, a composition with the
map $P$ will give us an involutive family of $Ad_{\P_+}$-invariant integrals
of the Toda flow. 

In our construction we utilize the idea of {\it the shift of argument}
that was introduced by Manakov [Ma] and then developed by  Mischenko
and Fomenko [MiFo2]. In these works the ``shifts'' $I_{\mu,a}(g)=I(g+\mu a)$
in the direction of a generic vector $a$ were used to construct a maximal
Poisson commutative family of functions on adjoint orbits  in semisimple
algebras. Two important distinctions in our case are that, firstly, functions
$I_{\mu,a}$ are not constant along the Toda flows and so we have first to apply
a nonlinear map $P$ to $g$, and, secondly, we use the shift of argument
in a ``maximally nongeneric'' direction of $e_m$ .







Namely, for an $Ad_\G$-invariant function $I\in C^\infty(\g)$ we define a 
function $I_\mu \in C^\infty (\P_+)$ by

$$ I_\mu (p) = I(\lambda e_{-m} + p + \mu e_m) \ . \eqno(7.1)$$

Denote
$$p_\mu= \lambda e_{-m} + p + \mu e_m = 
\lambda  e_{-m} + g' + v + (x_+ + \mu) e_m \ .$$
Then the equality
$$ [ \nabla I (p_\mu), p_\mu] = 0  \eqno(7.2)$$
and (2.3), (2.6), (2.7) imply
$$\eqalignno
{& \langle \nabla I (p_\mu), h_m\rangle  = 0\cr
& [\pi_{\V_+ } \nabla I (p_\mu),v] = 0 \ . &(7.3)\cr}
$$
Note also that
$$ \nabla I_\mu (p) = \pi_{\P_-} \nabla I (p_\mu) \ .\eqno(7.4)$$

Let $\{ \ , \ \}^\lambda$ be the Poisson bracket defined by (4.7), (4.8).

\proclaim{Theorem 7.1} Let $I,J$ be $Ad_G$-invariant functions. Then for any 
$\mu$ and $\nu$
$$\{ I_\mu , J_\nu\}^\lambda = 0 \ . \eqno(7.5)$$
\endproclaim

\demo{Proof} First, consider $\{ I_\mu , J_\nu\}_0$. Since, by (7.2), (7.3)
$$\eqalignno
{[p, \pi_{\P_-} \nabla I (p_\mu)]&= [p, \nabla I (p_\mu)]-
[p, \pi_{\V_+ } \nabla I (p_\mu)]\cr
&= - \mu [ e_m, \nabla I (p_\mu)] -
\lambda [ e_{-m}, \nabla I (p_\mu)] - 
[g', \pi_{\V_+ } \nabla I (p_\mu)]\ , &(7.6)\cr}
$$
we have
$$\eqalignno
{\{ I_\mu , J_\nu\}_0 (p)&= 
\langle [p, \pi_{\P_-} \nabla I (p_\mu)], \pi_{\P_-} \nabla 
J_\nu(p_\nu)\rangle \cr
&= - \mu \langle[ e_m, \nabla I (p_\mu)], \pi_{\P_-} \nabla 
J_\nu(p_\nu)\rangle 
+ \langle [\pi_{\V_+ } \nabla I (p_\mu), g'], \pi_{\P_-} \nabla 
J_\nu(p_\nu)\rangle \cr
&= -\mu \langle e_m, [\nabla I (p_\mu),\nabla J_\nu(p_\nu) ]\rangle  +
\langle \pi_{\V_+ } \nabla I (p_\mu), [g', \pi_{\P_-} \nabla 
J_\nu(p_\nu)]\rangle \cr
&= -\mu \langle e_m, [\nabla I (p_\mu),\nabla J_\nu(p_\nu) ]\rangle 
+ \langle \pi_{\V_+ } \nabla I (p_\mu), [g', \nabla J_\nu(p_\nu)]\rangle 
 \ .\cr}
$$
But, due to (7.2)
$$ [g', \nabla J_\nu(p_\nu)]= - [\nu e_m + v, \nabla J_\nu(p_\nu)] -
\lambda [e_{-m}, \nabla J_\nu(p_\nu)] \ $$
and by (2.3), (7.3)
$$ [\pi_{\V_+ } \nabla I (p_\mu), \nu e_m + v] = 0 \ .$$
Therefore,
$$\eqalignno{
\{ I_\mu , J_\nu\}_0 (p)&= -\mu \langle e_m, [\nabla I (p_\mu),\nabla 
J_\nu(p_\nu) ]\rangle 
- \lambda \langle \pi_{\V_+ } \nabla I (p_\mu), [e_{-m}, \nabla 
J_\nu(p_\nu)]\rangle \cr
&= \langle - \mu e_m + \lambda e_{-m}, [\nabla I (p_\mu),\nabla J_\nu(p_\nu) 
]\rangle \ . 
\cr}
$$
On the other hand, similar considerations show that

$$\{ I_\mu , J_\nu\}_0 (p) =  
\langle - \nu e_m + \lambda e_{-m}, [\nabla I (p_\mu),\nabla J_\nu(p_\nu) 
]\rangle \ .$$
Comparing the last two equations we obtain
$$\langle e_m, [\nabla I (p_\mu),\nabla J_\nu(p_\nu) ]\rangle =0 $$
and
$$\{ I_\mu , J_\nu\}_0 (p) = 
\lambda \langle e_{-m}, [\nabla I (p_\mu),\nabla J_\nu(p_\nu) ]\rangle \ 
.\eqno(7.7)$$

Now consider $\{ I_\mu , J_\nu\}_1$. Due to (7.4), (7.6)
$$ \eqalignno{
\pi_{\V_-} [ \nabla I_\mu (p), p] &= \lambda [ e_{-m}, \nabla I (p_\mu)]\cr
\pi_{\V_-} [ \nabla J_\nu (p), p] &= \lambda [ e_{-m}, \nabla J (p_\nu)]\ 
.\cr}
$$
Therefore,
$$\eqalignno{
\{ I_\mu , J_\nu\}_1 (p)&= 
\lambda^2 \langle [e_m, [ e_{-m}, \nabla I (p_\mu)]],[ e_{-m}, \nabla J 
(p_\nu)]\rangle \cr
&= \lambda^2 \langle\nabla I (p_\mu), [ e_{-m}, \nabla J (p_\nu)]\rangle \cr
&=- \lambda^2 \langle e_{-m}, [\nabla I (p_\mu), \nabla J (p_\nu)]\rangle   \ 
.&(7.8)\cr}
$$

Finally, it follows form (7.7), (7.8) that
$$
\{ I_\mu , J_\nu\}^\lambda (p)=
\{ I_\mu , J_\nu\}_0 (p) + \lambda^{-1} \{ I_\mu , J_\nu\}_1 (p) = 0 \ .$$
\qed
\enddemo

To conclude this section, we prove the following

\proclaim{Lemma 7.1} Let $l_m(p)$ be the function defined by (5.6). Then

\offset {20pt}{(i)} $\{\ l_m ,\ f\}_0 = 0$  for any $f\in C^\infty (\P_+)$\ .

\offset {20pt}{(ii)} $\{\ l_m ,\ f\}_1 = 0$  for any $\G'$-invariant
$f\in C^\infty (\P_+)$\ .

\endproclaim

\demo{Proof} (i) follows from the equality $[e_{-m}, \nabla  f]=0$

To prove (ii), recall that if $f$ is $\G'$-invariant then
$\langle[\nabla  f,g'],p\rangle  = 0$ for any $g'\in \g'$.
Let $p=g'+v+ x_+ e_m$ . Then
$$\eqalignno
{&\langle ad_{e_m} \pi_{\V_-} [\nabla f (p), p] , \pi_{\V_-} [e_{-m}, p] 
\rangle 
= - \langle\pi_{\V_-} [\nabla f (p), p], [e_m,[e_{-m}, p]] \rangle \cr
&= \langle [p, \nabla f (p)], v\rangle  = \langle [p, \nabla f (p)], 
g'+v\rangle \cr
&=\langle \nabla f (p), [g'+v, p]\rangle  = \langle \nabla f (p), 
[p,p]\rangle  = 0 \  .\qed\cr}$$
\enddemo

\head 8. Independence
\endhead

Let $I_1,\dots , I_r$ be polynomials that generates the algebra
of $Ad_G$-invariant functions. Consider polynomial functions $I_{1\mu},
\dots , I_{r\mu}$ defined by (7.1). Then
$$I_{i\mu}(p) = I_i(p_\mu) = \sum_{j=0}^{n_i} I_{ij}(p) \mu^j \ .
\eqno(8.1)$$

By Theorem 7.1, functions $I_{ij}$ are in involution w.r.t. 
$\{\ , \ \}^\lambda$ .

Let $p=g'+v+ x_+ e_m$. We want to prove 

\proclaim{Theorem 8.1}
For generic fixed 
$g'$ and $ x_+$ we can find among $I_{ij}$ considered as functions
on $\V_+$ at least $N={1\over 2} \dim\V_+ $ almost everywhere independent f
unctions.
\endproclaim

Note first that it is sufficient to prove this theorem for
$ x_+=0$ and $g'=h$, where $h$ belongs to the Cartan subalgebra $\h'$
of $\g'$. Moreover, it suffices to show that there exist such
$v\in \V_+$ that
$$ \dim Span \{ \pi_{\V_-} \nabla I_{ij} (p)\} \ge N \ . \eqno(8.2)$$

>From now on we fix  $\lambda$ to be $1$. Recall, that $I_{1} (g) = \langle 
g,g\rangle $
and , in particular, $1/2 \nabla I_{1}(p_\mu)= p_\mu = e_m + p + \mu e_m$.
We can choose $c_i,\ i=2,..., r$ in such a way that
$$ \langle\nabla I_{i}(p_\mu) - c_i p_\mu, e_m\rangle  = 0 \ .  $$
Then 
$$ Y_i(p_\mu) = \nabla I_{i}(p_\mu) - c_i p_\mu\ , i=2,..., r $$
satisfy
$$ [Y_i(p_\mu), p_\mu] = 0 \ . \eqno(8.3)$$
Due to (8.1),
$$Y_i(p_\mu) = \sum_{j=0}^{n_i} X_{ij}(p) \mu^j \ , $$
where $X_{ij}(p)$ are polynomial $\g$-valued functions of $v$.
Then
$$\pi_{\V_-}\nabla I_{ij} (p)= \pi_{\V_-} X_{ij}(p)$$
and (7.2) is
equivalent to the system of equations (cf. [Fo], sec 7.5.1)
$$\eqalignno{
& [X_{i0} (p), e_{-m} + p] = 0 \cr
&  [X_{ij} (p), e_{-m} + p] = 
[e_m, X_{i,j-1} (p)] \ \ \ \ \ j=1,\dots,n_i \ & (8.4) \cr
& [e_m, X_{i,n_i} (p)] = 0 \cr}
$$
By (7.3), (8.3) we have
$$\eqalignno
{& \langle X_{ij} , h_m\rangle  = 0\cr
& \langle X_{ij} , e_m\rangle  = 0 \cr
& [\pi_{\V_+ } X_{ij}, v] =0\ . \cr}
$$
Also , it follows from (8.4) that
$$\eqalignno{
&  [\pi_{\V_+ }X_{ij} (p), e_{-m}] + [\pi_{\V_- }X_{ij} (p), h]= 0\ & (8.5) 
\cr
&  [\pi_{\V_+ }X_{ij} (p), h] + [\pi_{\g' }X_{ij} (p), v] = 
[e_m, \pi_{\V_- } X_{i,j-1} (p)]\cr}
$$

Note that first of the equations (8.5) can be rewritten as
$$ \pi_{\V_+ }X_{ij} (p) =  
- ad_{e_m} ad_h \pi_{\V_- }X_{ij}(p)  \ . \eqno(8.6)$$

Let us now represent each $X_{ij}(p)$ as a sum
$$X_{ij} (p)= \sum_k X^k_{ij}(p)\ , $$
where $X^k_{ij}(p)$ is a  homogeneous $\g$-valued polynomial of degree $k$.

\proclaim{Lemma 8.1}
$X^0_{ij}(p)=\pi_{\g' }X^0_{ij} (p)$ . Moreover,
 $X^0_{i0}(p)\ , \ i=2,\dots, r$ 
generate $\h'$ .
\endproclaim

\demo{Proof} 
It is not hard to see that for generic $h$ element $p_{0 \mu}=
e_{-m} + h + \mu e_m $ is a regular element of $\g$. Therefore,
$ \nabla I_{i}(p_{0\mu})\ , i=1, \dots, r $ span a stabilizer of
$p_{0 \mu}$ in $\g$ which coincides with 
$ \h' + \Bbb C \{e_{-m} +  \mu e_m \}$ . Then $Y_i (p_{0 \mu})$
span $ \h'$ . Moreover, since the algebra of $Ad_G$-invariant functions
can be generated by trace polynomial of the adjoint representation
and since $\operatorname{Tr} (ad^n_{e_{-m} + h})= 
\operatorname{Tr} (ad^n_{e_{-m}})$ , $X^0_{i0}(p)\ , \ i=2,\dots, r$
also span $ \h'$ .
\qed
\enddemo

Let us now consider 
$$ L_{ij} := \pi_{\V_- }X^1_{ij}(p) \ .\eqno(8.7)$$ 
Due to Lemma 8.1, (8.5) and (8.6),
$$ \eqalignno
{&ad_{e_m} ad^2_h L_{ij} = ad_{e_m} L_{i,j-1}
- [X^0_{ij}, v] \ , \ j=1,\dots, n_i\cr 
&ad_{e_m} ad^2_h L_{i0} = - [X^0_{i0}(p), v] \ . \cr}
$$
or
$$ \eqalignno
{ & L_{i0} = 
-  ad^{-2}_h [X^0_{i0}, v_-] \ , \cr
& L_{ij} = ad^{-2}_h \left ( L_{i,j-1} -
 [X^0_{ij}, v_-] \right )\ , \ j=1,\dots, n_i \ ,& (8.8)\cr }
$$
where $v_-= ad_{e_{-m}} v$ ( here we used the fact that $ X^0_{ij}(p) \in 
\h`$
and $ad_{e_{-m}}$ commutes with $ad_{\g'}$.)

We denote $[X^0_{ij}, v_-]$ by $w_{ij}$ and the restriction
of $ad^{-2}_h$ on $\V_-$ by $A$. Then (8.8) can be rewritten as
$$ L_{ij} = - \sum_{l=0}^j A^{l+1} w_{i,j-l} \ \ . \eqno(8.9) $$

It follows form Lemma 8.1 that $L_{i0}, \ i=2,\dots, r$ are independent
and each $w_{ij}, \ j \ge 1$ is a linear combination of 
$w_{i0}, \ i=2,\dots, r$ \. If $V_- = ad_{\h'} v_-$ then (8.9)
implies
$$ Span \{ L_{ij}\} = Span \{ A^l V_- \ , l \ge 1\} \ . \eqno(8.10)$$

Now we can finish 

\demo{Proof of Theorem 8.1}
Recall (Proposition 2.1) that each weight space of $\h'$ in $\V_-$ is 
spanned by the vector $e_\alpha\ ,
\alpha \in \Phi^-_\F, \ \alpha \ne -m $. The highest weight vector is
$e_{-\alpha_{i_0}}$ and the lowest weight vector is $e_{\alpha_{i_0}-m}$ , where
$\alpha_{i_0}$ is the only simple root connected with $-m$ in the extended
Dynkin diagram of $\g$. (As before, we assume that $\g$ is not of type $A_r$.)

Since $(m-\alpha)(h)=-\alpha (h)$ for any $h\in\h'$, 
$ad_{\h'}$ acts by diagonal 
matrices  skew-symmetric w.r.t. the antidiagonal. Therefore, for generic $h$,
$A$ acts by the diagonal matrix  symmetric w.r.t. the antidiagonal and
first $N$ diagonal entries of $A$ are nonzero and distinct. 

We may assume that all coordinates of $v_-$ w.r.t the weight decomposition
are nonzero. Then one can find vector $v'_-\in V_- = ad_{\h'} v_-$ 
with all coordinates nonzero.
Then the standard considerations involving Vandermonde determinants show
that $Span \{ A^l v'_- \ , l \ge 1\}$ is $N$-dimensional, which, 
in view of (8.7), (8.10) proves the theorem.
\qed
\enddemo

\head 9. Proof of Main Theorem
\endhead

We proceed by induction. Assume that for simple algebras of rank less
then $r$ we constructed the maximal Poisson commuting family of integrals
of the Toda flow on generic coadjoint orbits. 

Let $O_{\zeta_0}$ be a generic coadjoint orbit in 
$\epsilon + \b_{-} \subset \g$,
where $\zeta_0$ is a normal form of the orbit given by (3.3) and let
$f_1,\dots, f_n$ be functions on $\g'$ whose restrictions to generic
orbits in $\epsilon' + \b'_{-}$ provide a maximal family of integrals
for the Toda flows. For $\zeta=\epsilon + x_- e_{-m} + x_0 h_m + v_- + \zeta'
 \in O_{\zeta_0}$ , where $v_-\in\V_-$ and $\zeta'\in \b'_-$, consider 
$\phi_1(\zeta)$ . If $v_-=0$ then, by (3.2) and
the definition of the $1$-chop $\phi_1$, $\phi_1(\zeta)$ is equal
to $\pi_{\g'}\zeta$ and belongs to a generic orbit in $\epsilon' + \b'_{-}$ .
This ensures that functions $f_1\circ\phi_1,\dots, f_n\circ\phi_1$ are almost
everywhere independent on $O_{\zeta_0}$ . 

Due to Proposition 2.2 and Theorem 4.1
$$\{f_i\circ\phi_1, f_j\circ\phi_1\}_{\epsilon + \b_-}= 0 \ ,$$
while Theorem 7.1, Proposition 4.1 and (5.5) imply that
$$\{I_{ij}\circ P, f_k\circ\phi_1\}_{\epsilon + \b_-}= 0 \ ,$$
where functions $I_{ij}$ are given by (8.1). Futhermore, by Lemma 7.1,
the function $l_m\circ P$, where $l_m$ is defined by (5.6) is in involution
with all $I_{ij}\circ P$ and $f_k\circ\phi_1$.

Let $I_{i_qj_q}, q=1,\dots,N$ be independent over $\V_+$ functions
among $I_{ij}$ whose existence is guaranteed by Theorem 8.1. 
Remark 3 of section 3.1 shows that functions 
$l_m\circ P;\ I_{i_qj_q}\circ P, q=1,\dots,N;\ 
f_k\circ\phi_1, k=1,\dots,n$ are independent on $\epsilon + \b_-$ .
Ivoking Proposition 3.3, we see that the number of functions in this family
is equal to ${1\over 2} \dim O_{\zeta_0}$ which concludes the proof of 
Theorem 1.1. \qed

\Refs

\offset{35pt}{[A]} Arhangelskij, A. A.:  Completely integrable Hamiltonian 
systems on the group of triangular matrices.  Mat. Sb. {\bf 108}, 134--142 
(1979)

\offset{35pt}{[BG]} Bloch, A. M. and Gekhtman, M. I. :
Hamiltonian and gradient structures in the 
Toda flows. In preparation.

\offset{35pt}{[DLNT]} P. Deift, L. Li, C. Nanda, C. Tomei: 
The Toda flow on
a generic orbit is integrable.  Comm. Pure \& Appl. Math., {\bf 39,}
183--232 (1986)

\offset{35pt}{[EFS]} Ercolani, N. M. Flaschka, H. , Singer, S. F.:
The geometry of the full Kostant-Toda lattice. Progr.  Math., {\bf 115,} 
181--225 (1993)

\offset{35pt}{[F]} Flaschka, H.: The Toda lattice I. Phys. Rev. B {\bf 9},
1924-1925 (1974)

\offset{35pt}{[Fo]}Fomenko, A. T.: Symplectic geometry. 
Advanced Studies in Contemporary Mathematics, 5. New York: 
Gordon and Breach Science
Publishers  1988

\offset{35pt}{[H]} Humphreys, J. E.:
Introduction to Lie Algebras and Representation
Theory. Berlin Heidelberg New York: Springer 1980

\offset{35pt}{[J]} Joseph, A.:
A preparation theorem for the prime spectrum of a semisimple Lie algebra.
J. Algebra {\bf 48}, 241--289 (1977)

\offset{35pt}{[K1]} Kostant, B.:\ 
The solution to a generalized
Toda lattice and representation theory.
Adv. Math. {\bf 34},
195--338 (1979)

\offset{35pt}{[K2]} Kostant, B.:\ 
The principal three-dimensional subgroup and the 
Betti numbers of a complex simple Lie groups.
Amer. J. Math. {\bf 81}, 973--1032 (1959)

\offset{35pt}{[K3]} Kostant, B.:\ 
Lie group representations on polynomial rings.
Amer. J. Math. {\bf 85}, 327--404 (1963)

\offset{35pt}{[LW]} Lipsman, R. L. and Wolf, J. A.:
Canonical semi-invariants and the Plancherel formula for parabolic groups.
Trans. AMS {\bf 269}, 111--131 (1982)

\offset{35pt}{[Ma]} Manakov, S. V.: Note on the integration of 
Euler's equations of the dynamics of an $n$-dimensional rigid body.
Funct. Anal. Appl. {\bf 10}, 328--329 (1977)

\offset{35pt}{[MiFo1]} Mischenko, A. S. and Fomenko, A. T.: 
A generalized Liouville
method for the integration of Hamiltonian systems.
Functional Analysis and its Applications {\bf 12}, 113--131 (1978)

\offset{35pt}{[MiFo2]} Mischenko, A. S. and Fomenko, A. T.: 
Euler equation on finite-dimensional Lie
groups.  Izvestija {\bf 12}, 371--389 (1978)

\offset{35pt}{[Mo]} Moser, J.:\ 
Three integrable Hamiltonian
systems connected with isospectral deformation.
Adv. Math. {\bf 16,}
197--220 (1975)

\offset{35pt}{[N]} Nehoroshev, N. N. :  Action-angle variables and their
generalizations. Trans. Moscow Math. Soc. {\bf 26}, 180--197 (1972)  

\offset{35pt}{[OP]} Ol'shanetsky, M.A. and Perelomov, A.M. : \ 
Explicit solutions of classical generalized Toda models.
Invent. Math. {\bf 54,} 261--269 (1979) 

\offset{35pt}{[OV]} Onishchik, A. L. and Vinberg, E. B. :
Lie Groups and Algebraic Groups. 
Berlin Heidelberg New York: Springer 1990

\offset{35pt}{[RSTS1]} Reyman A.G., Semenov-Tian-Shansky, M. A.: 
Reduction of Hamiltonian systems, affine Lie algebras and Lax equations I. 
Invent. Math. {\bf 54},
81-100 (1979)

\offset{35pt}{[RSTS2]} Reyman A.G., Semenov-Tian-Shansky, M. A.:
Group-theoretical methods in the theory of 
finite-dimensional integrable systems, Dynamical Systems VII, 
Encyclopaedia of Mathematical Sciences {\bf 16},
Berlin Heidelberg New York: Springer 1994

\offset{35pt}{[S1]} Singer, S.: Ph. D. Dissertation.
Courant Institute, 1991.

\offset{35pt}{[S2]} Singer, S.: Some maps from the full Toda lattice
are Poisson. Phys. Lett. A {\bf 174}, 66-70 (1993)

\offset{35pt}{[Sy]} Symes, W.W.: 
Systems of Toda type, inverse spectral problems, and representation theory.  
Invent. Math. {\bf 59,} 13--51 (1980) 

\offset{35pt}{[T]} Thimm, A.: 
Integrable Hamiltonian systems on homogeneous spaces.
Ergod. Th. and Dynam. Systems  {\bf 1}, 495--517 (1981)

\offset{35pt}{[Tr1]} Trofimov, V. V.: Euler equations on Borel
subalgebras of semisimple Lie algebras. Izvestija {\bf 43},
714--732 (1979)

\offset{35pt}{[Tr2]} Trofimov, V. V.: Finite-dimensional representations
of Lie algebras and completely integrable systems.
Mat. Sb. {\bf 11}, 610--621 (1980)

\end